# THE EFFECT OF GRAIN SIZE DISTRIBUTION ON THE DEFECT GENERATION MECHANISM OF 0201 PASSIVES

Swagatika Patra, Christopher. M. Greene, Ph.D., Daryl. L. Santos, Ph.D.
Watson Institute of Systems Excellence, Binghamton University-State University of New York
Binghamton, New York, USA
spatra1@binghamton.edu

Ganesh Pandiarajan, Ph.D., Satyanarayan "Satya" Iyer, Ph.D., Shawn Eckel, Narayanan Manickam
SMART Modular Technologies
Newark, California, USA

**ABSTRACT**

The recent advancements in electronics manufacturing has necessitated the demand for miniaturization of electronic components. In particular, the product sizes for passive components have evolved from 1005 (1.0x0.5mm) which is roughly the size of a grain of sand to 0201 (0.25x0.125mm) which is 1/16$^{th}$ the size of a grain of sand. Grain size distribution and composition play a vital role in tailoring the demands for reliably shrinking dimension of passives. The study focuses on analyzing the effect of grain size and composition, which plays a significant role in maintaining high yield output during the manufacturing processes. The objective of this study is to compare the grain morphology and its effect on defect generation mechanism for 0201 resistors. The experimental setup was prepared to evaluate the grain structure as per ASTM E112-3 test standards. The 0201 resistor samples from three different vendors were assembled on test vehicle PCB (Printed Circuit Board). The resulting observations of the study showed that tombstoning defect was observed during the reflow process for Vendor A. Scanning Electron Microscope (SEM) results showed the presence of microstructural difference in the tin grain size of resistors. The tin grain size for Vendor A was observed to be 2 microns, whereas the grain size for Vendors B and C were 4 and 5 microns, respectively. Furthermore, the results of grain size measurements performed using the intercept method showed that a grain size number of 13 and 13.6 was observed for Vendors B and C, whereas a grain size number of 15 was observed for Vendor A. The second phase of the study focuses on techniques for improving the grain size number to mitigate the tombstoning defect. It was observed that using an alternate plating solution during the manufacturing of passives resulted in decreased grain size number while containing the tombstoning issue.

**INTRODUCTION:**

With the recent advancements in miniaturization of electronic components, it is observed that the dimensions of microprocessor have been shrinking down per Moore's law. The dimensions of passive electrical components are also scaling down at an unprecedented pace with the decreasing size of the microprocessors [1]. The dimensions for passive components have evolved from 1005m (1.0x0.5mm) which is roughly the size of a grain of sand to 0201m (0.25x0.125mm) which is 1/16$^{th}$ the size of a grain of sand. The size scale for the 0201 and 01005 passive components ranges from the order of 100–300 μm. Millions of passive components are mounted onto the printed circuit board assembly (PCBA) every day across a wide range of products in the electronics manufacturing industry. Thus, it becomes critical to understand the failure mechanism of these components to maintain high manufacturing yields and long-term reliability of the product [1]. Tombstoning is one of the commonly occurring defects during the soldering process that involves lifting of one end of a passive component. The occurrence of tombstoning defect of passive component after the reflow process is shown in Figure 1. There are various factors affecting the tombstoning defect such as the reflow profile, component dimension and geometry, pad size, board design, pick and place accuracy, and varying grain morphology of components [2]. This paper focuses on the effect of varying grain morphology and size of 0201 passive resistors resulting in tombstoning defects.

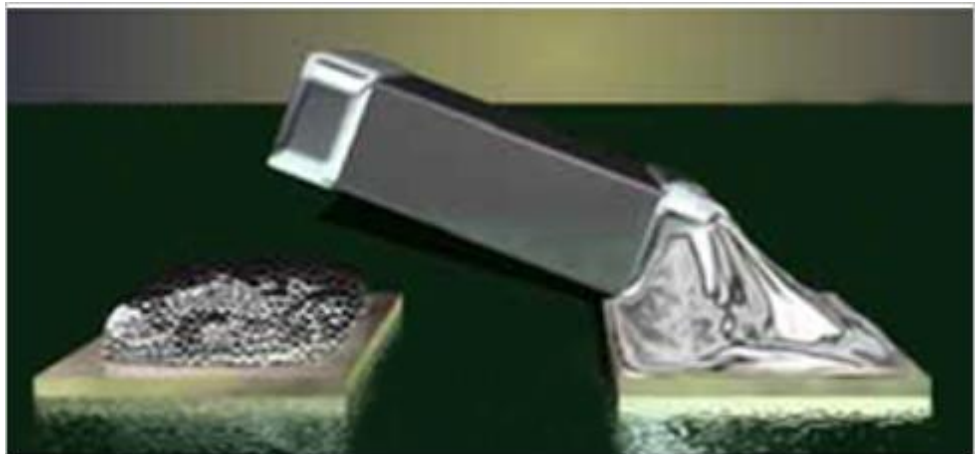
**Figure 1.** Tombstoning of passive component after reflow process

**Construction of Thick Film Resistor**
Prior to understanding the failure mechanism of the electrical passives, it is vital to understand the construction of the components. Resistors are passive electronic devices, which regulate the flow of electric current without using an external power source to function in a circuit [3]. There are different types of resistors varying from conventional ceramic elements, embedded resistors inside of a printed circuit board, thin film resistors, or thick film resistors. Figure 2 shows the schematic construction of a thick film resistor. Thick film resistors are commonly used in high performance electronic products and hybrid circuits for sensing current and power conversion [3]. These varieties of resistors are typically prepared by depositing a mixture of metal and particles onto a ceramic substrate. It is then subjected to a high temperature (typically 850°C or so in air) which produces a conductive cermet matrix. After laser trimming to value, a coating of a layer of glass insulator is applied on the top for protection against environment [4]. Thus, given the intricacy of the construction of these films, it is critical to understand and access the dependence of the physical properties of film, namely the thermal, electrical, and optical properties on the grain structure analysis. The microstructural grain size analysis is critical to understand the evolution of grain growth and its effect on the resulting defect generation mechanism intrinsic to the electrical component in use [5].

**Importance of Grain Morphology Analysis:**
The construction of high-quality thick film passives depends on two sets of parameters: sputtering process parameters and post-deposition treatment parameters. The sputtering process parameters include RF power, gas ambient flow (oxygen, argon, hydrogen, and H2O vapor) flow rate, sputtering working pressure, deposition rate, substrate temperature, substrate type, target to substrate distance, and target density used [3]. The post deposition treatment parameters include post annealing, ionization irradiation, laser irradiation and interlayers on indium-tin oxide films [6]. In addition to these parameters, thickness and the plating solution also influences the properties of the electronic components. Electrical and thermal properties of the thick film passives depends on the physical properties such as the crystal structure and oxidation level of the components. Hence, an appropriate optimization of these properties require a fundamental understanding of the elemental composition and evolution of grain growth. For example, the grain size of resistors films are affected by film thickness and annealing temperature and the plating chemistry used during the manufacturing process [7]. Thus, this paper attempts to investigate the effect of grain size on the tombstoning of 0201 passive resistors.

The control of grain size is vital for understanding the electrical and mechanical properties of metals. The restricted dislocation motion of electrons is associated with fine grain size and defined grain boundary, which increases strength of the materials. As defined by the Hall-Petch equation, the relationship between yield strength ($\sigma_0$) and grain size ($D_g$) is given by
$\sigma_0 = \sigma_s + BD_g^{-1/2}$
where $\sigma_s$ is the yield stress of a single crystal and *B* is a constant. The relationship per the equation states that there is an increase in the stress with decreasing grain size. Hence, a larger and uniform grain size and morphology is desired for better performance of the electronic components [8].

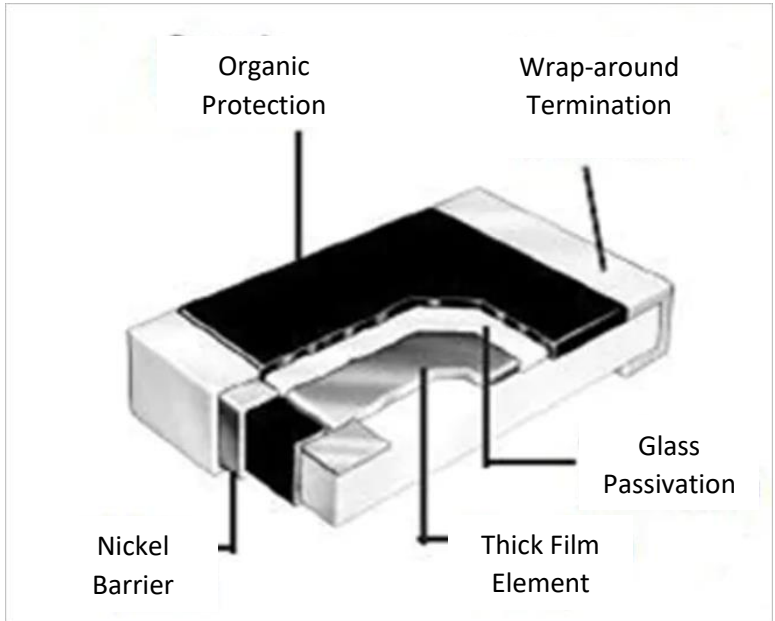

**Figure 2.** Construction of thick film resistor chip [4]

**EXPERIMENTAL SETUP:**
The experimental setup to understand the effect of grain size on the tombstoning defect was performed using thick film anti-sulfurated 15 Ohm 0201 chip resistors. The 0201 resistor samples from three different vendors were assembled on test vehicle PCBs. The 15 ohm 0201 resistors were mounted onto the PCB using 5.8LS solder paste. The PCBs were then reflowed to validate the mechanism of defect generation. The evaluation of the grain structure was performed as per ASTM E112-3 test standards using the intercept method. The intercept method was used evaluate the actual count of grains intercepted and grain boundary per unit length of test line. The intercept method was used to represent the actual count of the number of grain boundary intersections with a test line, per unit length, to calculate the mean lineal intercept length. The calculated length was used to determine the ASTM grain size number (G). The calculation of grain size using the intercept method was performed using the Leica Application Suite (LAS) software. The precision of this method is defined as a function of the number of intercepts or intersections counted. SEM analysis was performed to understand the microstructural difference between the samples from vendors. The experimental design setup is outlined below in Table 1.

Table 1. Experimental design setup

| Vendor | Sample Size | Number of Resistor Locations per Board |
|---|---|---|
| Vendor A | 3 boards | 20 |
| Vendor B | 3 boards | 20 |
| Vendor C | 3 boards | 20 |

## RESULTS AND DISCUSSION:

There was no occurrence of tombstoning defect in case of Vendor C. The yield percentage based on the samples taken for the experimental study is summarized in Table 2. A comparative yield analysis was performed based on the tombstoning defect observed for all three vendors in Figure3.The results of the study showed that high percentage of tombstoning defect was observed in case of Vendor A (shown in Figure 4).

Table 2. Yield summary of experimental results

| Vendor | Sample Size | Number of Resistor Locations per Board | Number of Tombstoning Defect Observed per board | Total number of Tombstoning Defect Observed | Yield |
|---|---|---|---|---|---|
| Vendor A | 3 boards | 20 | Board 1: 8 | 20 | 66.6% |
| | | | Board 2: 5 | | |
| | | | Board 3: 7 | | |
| Vendor B | 3 boards | 20 | Board 1: 0 | 1 | 98.3% |
| | | | Board 2: 1 | | |
| | | | Board 3: 0 | | |
| Vendor C | 3 boards | 20 | Board 1: 0 | 0 | 100% |
| | | | Board 2: 0 | | |
| | | | Board 3: 0 | | |

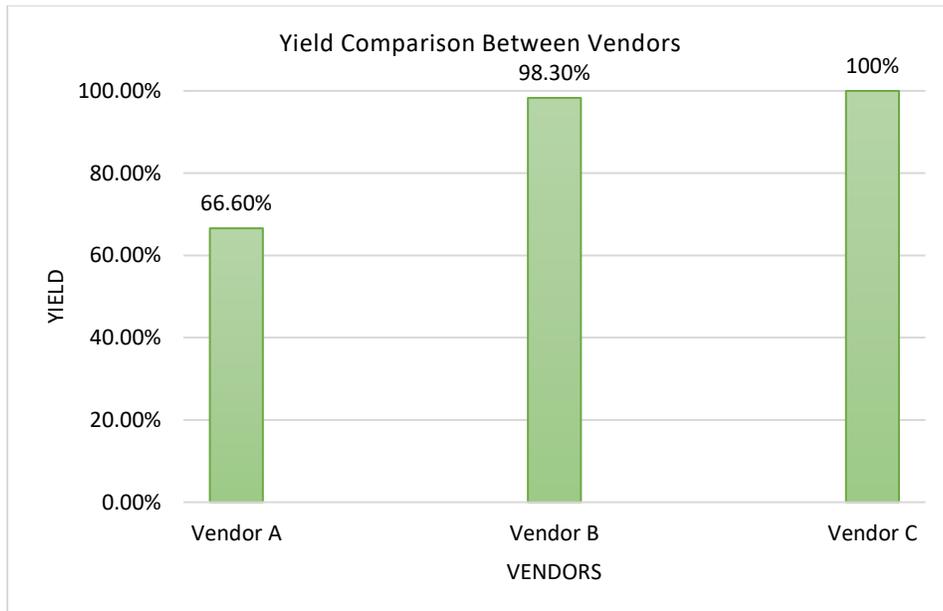

Figure 3. Comparative yield analysis for Vendor A, B and C

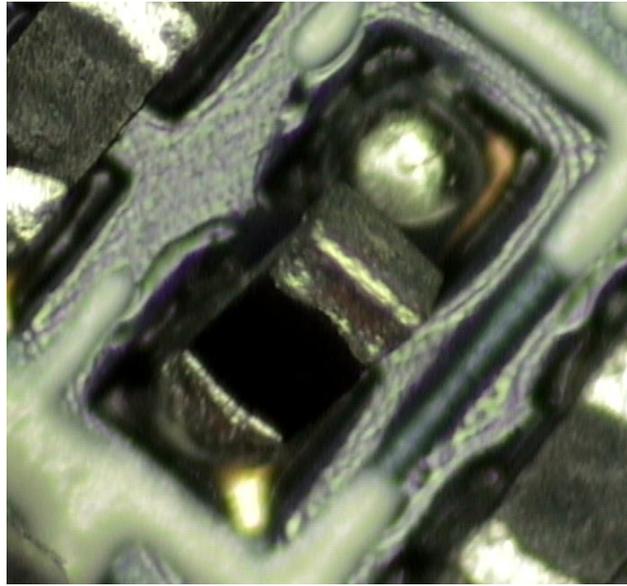

Figure 4. Tombstoning of 15 Ohm 0201 resistor on samples from Vendor A

The root cause analysis to evaluate the cause of tombstoning defect was analyzed using the following steps:

i. **Grain Size Analysis using SEM**
The results of the SEM analysis performed on the samples from Vendors A, B and C showed microstructural difference in the tin grain size of the components. The point of analysis of the resistor samples is shown in Figure 5.

A tin grain size of 2µm was observed in case of Vendor A, whereas in case of Vendor C, a grain size of 5µm was observed. The comparative SEM analysis showing the microstructural difference in the tin grain size between vendors is shown in Figure 6 and Figure 7. Furthermore, there was no distinct grain boundary lines observed in case of Vendor A, resulting in dislocation along the lines, thus making the component prone to failure. In addition to this, as indicated by past studies [7], the electrical and the thermal conductivity of the material tends to decrease with decreasing grain size.

The results of the experiment were in accordance with the observation made from the SEM analysis and the past research studies. The study performed showed that the 0201 passive components with smaller tin grain size (2 µm) resulted in higher number of tombstoning defects as compared to the samples with higher tin grain size. Thus, it is recommended that the tin grain size should be larger and uniform in order to mitigate the tombstoning issue.

Table 3. Result Summary for Grain Size Analysis

| Vendor | Grain Size | Total number of tombstoning defect observed |
|---|---|---|
| Vendor A | 2µm | 20 |
| Vendor B | 4µm | 1 |
| Vendor C | 5µm | 0 |

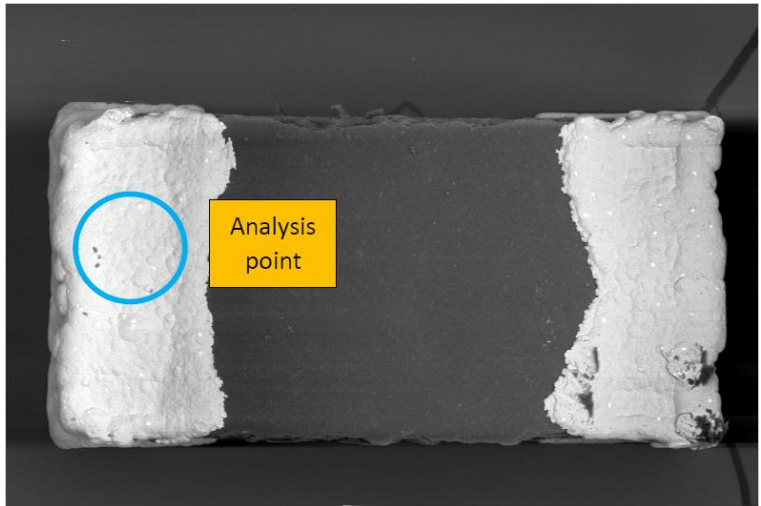
Figure 5. Point of analysis for evaluating the grain size

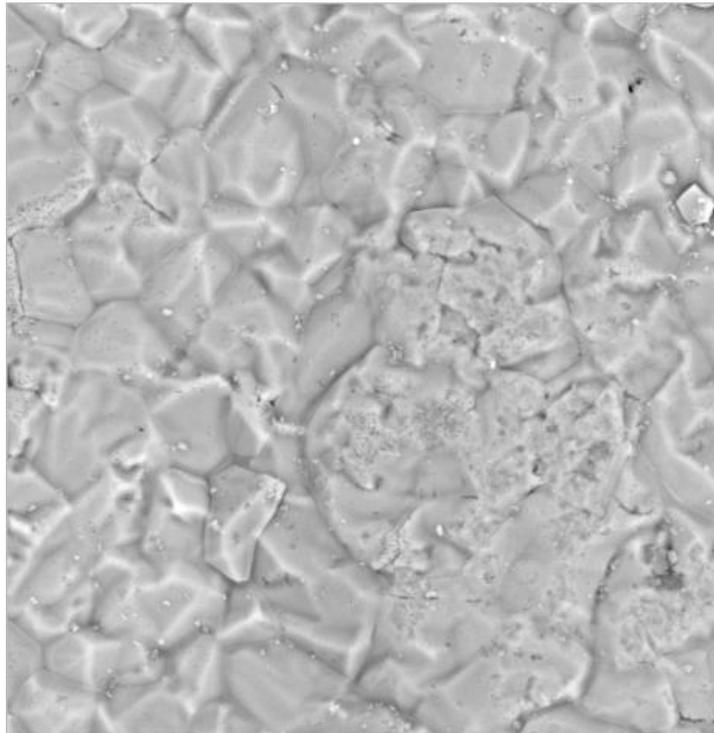
Figure 6. SEM analysis of samples from Vendor A showing non-uniform grain morphology

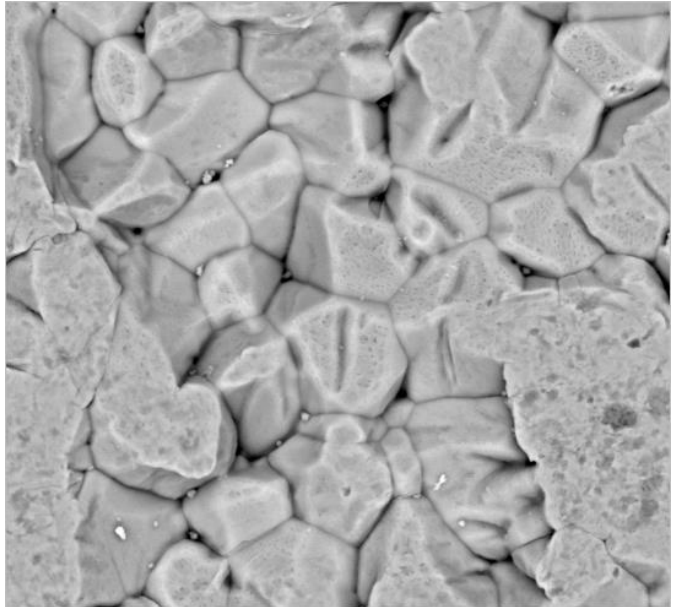

Figure 7. SEM analysis of samples from Vendor C showing uniform and larger grain size with distinct grain boundaries

**ii. Evaluation of Grain Size Number**

ASTM grain size number (n) is related with the number of grains that can be counted in 100X magnification (N) as defined by the equation,

$$N=2^{(n-1)}$$

Therefore, per the above relation, ASTM grain size number increases with decreasing grain size [9]. The results of the grain size number showed that lower grain size number of 13 was observed in the case of Vendor C. On the contrary, the highest grain size number of 15 was observed in the case of Vendor A. Hence, this shows that per the above definition, the grain size number decreased with an increase in the grain size. A comparative analysis of difference in the 0201 resistor between 3 different vendors is shown in Table 4. The table shows that there is a difference in the tin grain size and the thickness between the vendors. This affects the defect generation mechanism resulting in tombstoning of the component during the reflow process.

Table 4. Comparative analysis in the construction of the components between different vendors

| Description | Vendor A | Vendor B | Vendor C |
|---|---|---|---|
| Length (mm) | 0.607 | 0.578 | 0.598 |
| Width (mm) | 0.297 | 0.291 | 0.291 |
| Height (mm) | 0.212 | 0.223 | 0.223 |
| Bottom Electrode Thickness (mm) | 0.156 | 0.15 | 0.15 |
| Electrode (Sn) Thickness (µm) | 6.5 | 8.3 | 9.1 |
| Tin Plating Layer -Tin grain size (µm) | 2 | 4 | 5 |

In the second phase of the study, attempts were made to increase the tin grain size of Vendor A. This analysis was performed to validate if enhancing the tin grain size of Vendor A can help to combat the tombstoning issue. The plating solution was changed from acidic to neutral solution. The annealing temperature was also increased to enhance the grain size of the component. As specified by research study in [10] the size of grain crystallites is directly related to annealing temperature, i.e., the size of grains increases with an increase in temperature. As the annealing temperature was increased, the microstructure of the grains was observed to become larger in size and uniform in morphology. The results of the experiment using the new samples from Vendor A showed that there was no occurrence of tombstoning defect. The results of the study are summarized in Table 5. The experiment to validate the new samples from Vendor A was performed using a sample size of 3 PCBs with 15 Ohm 0201 resistors mounted onto 20 locations across each board.

It was also observed that changing the plating solution and annealing temperature during the processing of the components helped to increase the tin grain size and the grain boundary of the components. The results were confirmed after performing SEM analysis to validate the grain size and distribution and evaluating the grain size number as per ASTM E112. SEM analysis of the grain size is shown in Figure 8.

Table 5: Summary of Results for Vendor A

| Vendor | Grain Size | Grain Size Number | Total number of tombstoning defect observed | Electrode (Sn) Thickness (µm) |
|---|---|---|---|---|
| Vendor A | 4.5 µm | 13.8 | 0 | 8.7 |

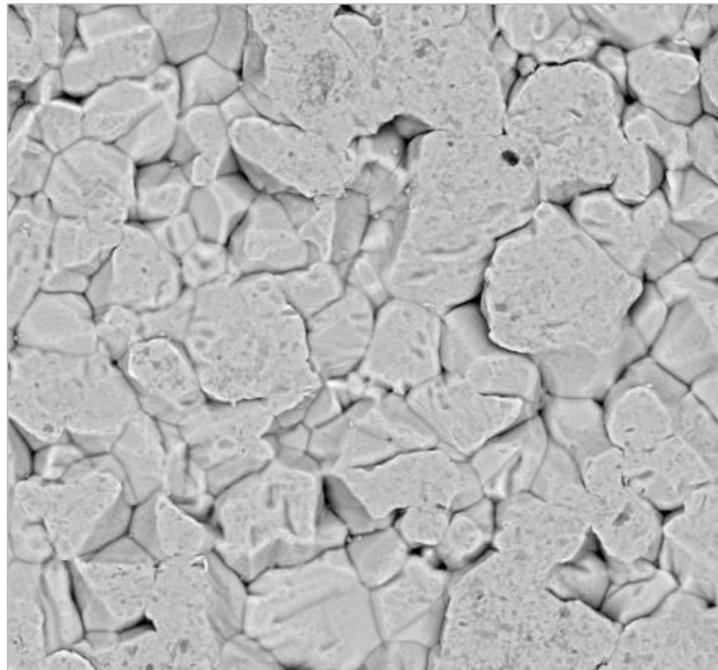

Figure 8. Improved grain size and boundary of Vendor A after using alternate plating solution and annealing temperature

## CONCLUSION:

The results of the study showed that grain size and morphology plays a significant role in the defect generation mechanism of the 0201 passive components. The experimental analysis showed that the electronic components with tin grain size of 2 µm showed the highest number of tombstoning defects, whereas the components with grain size number of 5 µm did not show any tombstoning defect. Furthermore, the results of the study showed that using of neutral solution for plating and increasing the annealing temperature showed an increase in grain size of the component from 2 µm to 5 µm. Also, it was observed that increasing the tin grain size helped to combat the tombstoning defect of the component.

## ACKNOWLEDGEMENTS:

The authors would also like to acknowledge the support of Watson Institute of Systems Excellence, Binghamton University and SMART Modular Technologies in this study, in particular Dr. Krishnaswami "Hari" Srihari and Dr. Mohammed Khasawneh. The authors would also like to thank Viking Tech Corporation for all the technical support and inputs during the experimental study.